\documentclass[aps,prd,twocolumn,showpacs,nofootinbib,amsmath,amssymb,floatfix,superscriptaddress,showkeys]{revtex4}
\usepackage{graphicx}
\usepackage{dcolumn}
\usepackage{bm}
\usepackage{captcont}
\usepackage{xcolor}
\usepackage{threeparttable}

\usepackage{epstopdf}
\usepackage{mathtools}
\usepackage{natbib}
\usepackage{ulem}
\usepackage{color,txfonts}
\definecolor{blue0}{rgb}{0,0,0.6}
\usepackage[colorlinks,linkcolor=blue0,anchorcolor=blue0,citecolor=blue0,urlcolor=blue0]{hyperref}

\newcommand{\jcap}{J. Cosmol. Astropart. Phys.}
\newcommand{\apjl}{Astrophys. J. Lett.}

\newcommand{\aap}{Astron. Astrophys}
\newcommand{\aj}{Astron. J.}

\newcommand{\nar} {NewA Rev.}

\newcommand{\hess}{H.E.S.S.}

\begin{document}
\title{Searching for the possible signal of the photon-axionlike particle oscillation in the combined GeV and TeV spectra of supernova remnants}
\author{Zi-Qing Xia}
\affiliation{Key Laboratory of Dark Matter and Space Astronomy, Purple Mountain Observatory, Chinese Academy of Sciences, Nanjing 210008, China}
\author{Yun-Feng Liang}
\email{Corresponding author. liang-yf@foxmail.com}
\affiliation{Key Laboratory of Dark Matter and Space Astronomy, Purple Mountain Observatory, Chinese Academy of Sciences, Nanjing 210008, China}
\affiliation{Laboratory for Relativistic Astrophysics, Department of Physics, Guangxi University, Nanning 530004, China}
\author{Lei Feng}
\affiliation{Key Laboratory of Dark Matter and Space Astronomy, Purple Mountain Observatory, Chinese Academy of Sciences, Nanjing 210008, China}
\affiliation{College of Physics, Qingdao University, Qingdao 266071, China}
\author{Qiang Yuan}
\affiliation{Key Laboratory of Dark Matter and Space Astronomy, Purple Mountain Observatory, Chinese Academy of Sciences, Nanjing 210008, China}
\affiliation{School of Astronomy and Space Science, University of Science and Technology of China, Hefei, Anhui 230026, China}
\affiliation{Center for High Energy Physics, Peking University, Beijing 100871, China}
\author{Yi-Zhong Fan}
\email{Corresponding author. yzfan@pmo.ac.cn}
\affiliation{Key Laboratory of Dark Matter and Space Astronomy, Purple Mountain Observatory, Chinese Academy of Sciences, Nanjing 210008, China}
\affiliation{School of Astronomy and Space Science, University of Science and Technology of China, Hefei, Anhui 230026, China}
\author{Jian Wu}
\affiliation{Key Laboratory of Dark Matter and Space Astronomy, Purple Mountain Observatory, Chinese Academy of Sciences, Nanjing 210008, China}
\affiliation{School of Astronomy and Space Science, University of Science and Technology of China, Hefei, Anhui 230026, China}

\date{\today}

\begin{abstract}

The conversion between photons and axionlike particles (ALPs) in the Milky Way magnetic field could result in the detectable oscillation phenomena in $\gamma$-ray spectra of Galactic sources.
In this work, the GeV (Fermi-LAT) and TeV (MAGIC/VERITAS/\hess) data of three bright supernova remnants (SNRs, ie. IC443, W51C and W49B) have been adopted together to search 
such the oscillation effect. 
Different from our previous analysis of the sole Fermi-LAT data of IC443, we do not find any reliable signal for the photon-ALP oscillation in the joint broadband spectrum of each SNR.
The reason for the inconsistence is that in this work we use the latest revision (P8R3) of Fermi-LAT data, updated diffuse emission templates and the new version of the source catalog (4FGL), which lead to some modification of the GeV spectrum of IC443. 
Then we set constraints on ALP parameters based on the combined analysis of all the three sources. 
Though these constraints are somewhat weaker than limits from the CAST experiment and globular clusters, they are supportive of and complementary to these other results.

\end{abstract}
\pacs{95.35.+d, 95.85.Pw, 98.58.Mj}
\keywords{Dark matter$-$Gamma rays: general$-$ISM: supernova remnants}

\maketitle

\section{Introduction}
\label{sec1}
Many observations such as galaxy rotation curves, the gravitational lensing and the CMB power spectrum, have shown that the total mass of the universe cannot be explained by the ordinary matter in the standard model.
Dark matter (DM) has been proposed to explain the missing mass, which is believed to make up 26 precent of the total energy of the current universe \cite{2016A&A...594A..13P}.
Various hypothetical particles have been proposed to explain this invisible form of matter by theoretical physicists, such as weakly-interacting massive particles (WIMPs), axions, axion-like particles (ALPs), sterile neutrinos and gravitinos. ALPs, one kind of attractive cold DM candidates \cite{DMA,DMA2012}, share the same intriguing property with axions that these particles and photons can convert to each other in electromagnetic fields through the Primakoff process.
This interaction is closely related to the mass of the ALP ($m_{a}$), the coupling constant ($g_{a\gamma}$) and the electromagnetic field.
The coupling $g_{a\gamma}$ of the ALP is independent of its mass $m_{a}$, which is different from the axion.

Ongoing global efforts have been made to probe axions or ALPs by the Primakoff process\cite{exp2015,expph2013}. 
The laboratory experiments, OSQAR \cite{OSQAR2014} and ALPS \cite{ALPSI2010}, are based on the ``Light shining through a wall" (LSW) approach \cite{light2011} and find no excess of events.
The ADMX has searched for dark matter axions (or ALPs) from the local galactic dark matter halo and excluded optimistic axion models in the $1.9 - 3.53$ $\mu$eV and $4.9 - 6.2$ $\mu$eV ranges \cite{ADMX2012}.
Moreover, axions or ALPs originating in the Sun's core via the Primakoff conversion of plasma photons, can be converted into visible photons inside a telescope with a magnetic field, which is the basis of the Axion helioscopes such as the CAST \cite{CAST2017} and the IAXO \cite{IAXO2015}.
One of the most competitive upper bounds has been set by the CAST on the coupling ($g_{a\gamma}< 6.6 \times 10^{-11}\,{\rm GeV}^{-1}$) over a large mass range \cite{CAST2017}.

In addition to the aforementioned ground-based experiments, astronomical observations also contribute to searches for axions or ALPs.
The photon-axion coupling in stellar cores would influence the stellar evolution and population properties of the stars in globular clusters. These effects have been applied to derive constraints on ALP parameters \cite{1987PhRvD..36.2211R, 2008LNP...741...51R, 2013PhRvL.110f1101F, 2014PhRvL.113s1302A}. For example, with measurements of the $R$ parameter, i.e. the number ratio of stars in horizontal over the red giant branch, by using a sample of 39 galactic globular clusters, Ref. \cite{2014PhRvL.113s1302A} derived a strong upper bound on the photon-ALP coupling of $g_{a\gamma} < 6.6 \times 10^{-11}\,{\rm GeV^{-1}}$.

Besides, the conversion between $\gamma$-ray photons and ALPs in external magnetic fields could create energy-dependent modulations in spectra of $\gamma$-ray sources, which is so called the photon-ALP oscillation.
Such a spectral oscillation effect could be a smoking-gun signature for the existence of ALPs \cite{Hochmuth07,hooper07alp,Angelis08,Simet08,sc09,Meyer13} and has been widely used to search for ALP signals \cite{belikov11alp,hess13pks2155,reesman14alp,berenji16alp_ns,fermi16alp,meyer17sne,zc16alp,Majumdar17psr1,Majumdar17psr2,ALP2018,ic443alp,liang18alp}. 
The Fermi-LAT Collaboration have searched for the photon-ALP oscillation signal in the High Energy (HE) spectrum of NGC 1275 and excluded the coupling $g_{a\gamma}$ above $5 \times 10^{-11}\,{\rm GeV}^{-1}$ for ALP masses $0.5 - 5$ neV \cite{fermi16alp}.
The HE and Very High Energy (VHE) photons from PKS 2155-304 have also been analyzed separately by the H.E.S.S. Collaboration \cite{hess13pks2155} and Zhang {\it et al.} \cite{zc16alp} without significant signal of ALPs found.
The $\gamma$-ray spectra of bright pulsars \cite{Majumdar17psr1, ALP2018, Majumdar17psr2} and SNRs \cite{ic443alp} obtained from the Fermi-LAT data show intriguing indications for the photon-ALP oscillation, but the best-fit parameters are in tension with the limit from the CAST helioscope.
Besides, Liang {\it et al.} found that higher ALP mass region of the ALP parameter space can be probed by the H.E.S.S. observations of some galactic bright sources \cite{liang18alp}.

\begin{figure}
\centering
\includegraphics[width=0.52\textwidth]{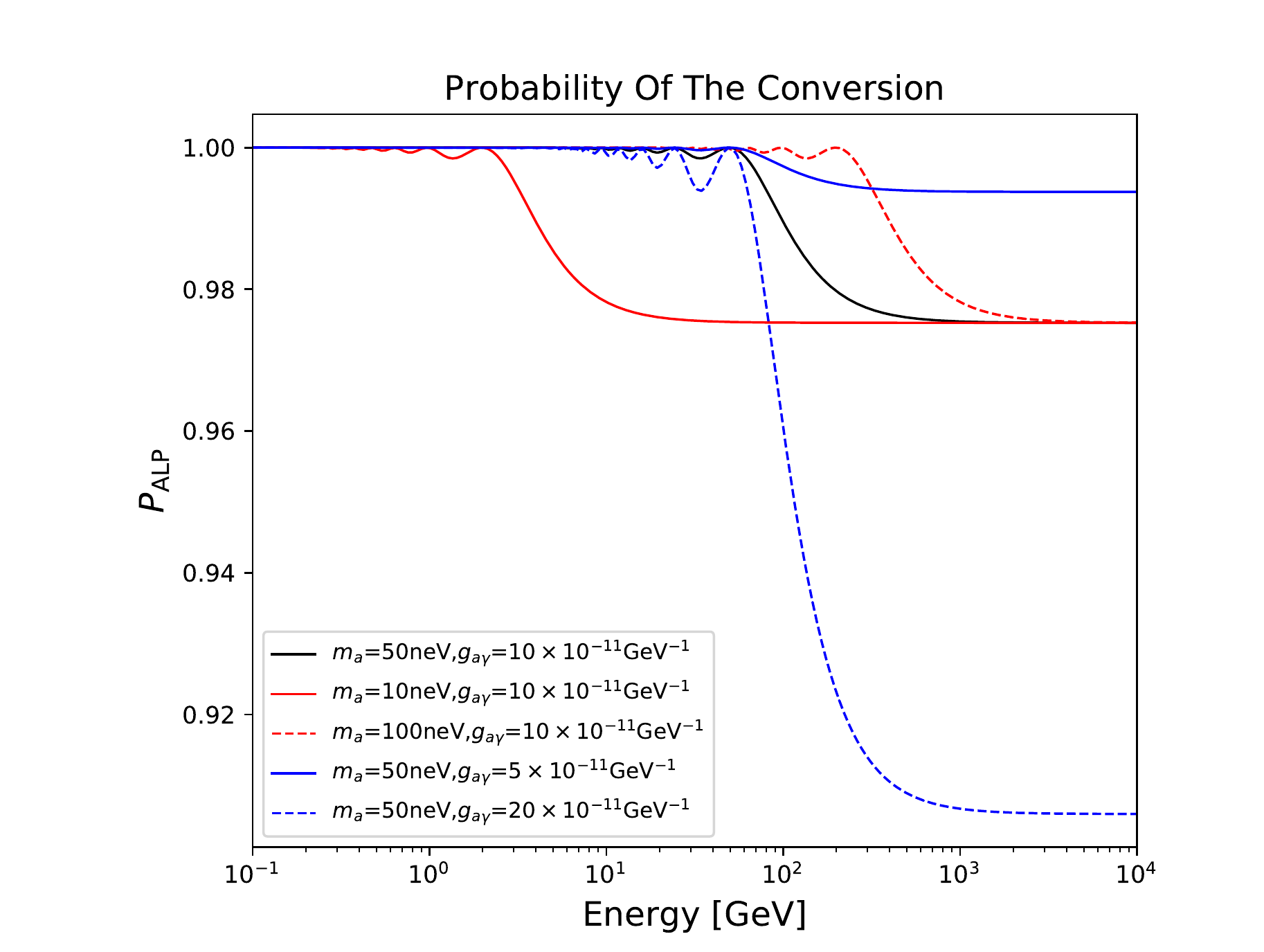}
\caption{Survival probabilities $P_{\rm ALP}$ for photons from IC443 with five different sets of parameters (Bfield1).}
\label{fig:PALP}
\end{figure}

The Figure1 in our early work \cite{ic443alp} indicates that the conversion probability of photons from IC443 with
the best-fit parameters increase to a high level (about 25\%) at energies above 10 GeV.
Imaging Atmospheric Cherenkov Telescopes (IACTs, MAGIC/VERITAS/\hess) have a high sensitivity in the VHE range from several tens of GeV to hundreds of TeV.
With the addition of IACTs observations, we will get broadband spectra from a few hundreds of MeV to 100 TeV, which would be very helpful in testing the intriguing indication found in the Fermi-LAT spectrum of IC443 \cite{ic443alp}.
Adopting these VHE data could also help to make more reliable constraints on high $m_a$, for which the oscillation would appear at TeV energies (see Figure \ref{fig:PALP} for illustration).

In this work, we combine the Fermi-LAT \cite{fermilat} data with the IACTs (MAGIC \cite{magic}, VERITAS \cite{veritas} and \hess \cite{hess}) measured spectra of three SNRs (IC443, W51C and W49B) to search for the possible imprint of the photon-ALP conversion and constraint on ALP parameters.

All the three SNRs in our sample have been established as strong stable $\gamma$-ray sources and have a relatively high-quality data both in the Fermi-LAT \cite{ic443fermi10, w51cfermi09, w49bfermi10} and IACTs\cite{ic443magic07, ic443veritas09, ic443veritas15, w51cmagic12, w49bhess18, 2HWC} observations. 
The stability of these sources are necessary for our joint-analysis since the VHE data and the Fermi-LAT data had not been collected simultaneously (If the source emission is instead viable, there is no reason to assume a constant spectrum).
As it will be shown below, the flux points of each SNR from the Fermi-LAT and Magic/\hess observations are found in good agreement, while a tension is found between the Fermi-LAT spectrum and the VERITAS spectrum for IC443.
In addition, each SNR is located in the Galactic plane, where magnetic fields along the lines of sight are relatively high.
The location and distance \cite{dSNR} information of these SNRs is listed in Table. \ref{tb1}.

\begin{table}[t]
\caption{The basic information and results}
\begin{ruledtabular}
\begin{center}
\begin{tabular}{ccccc}
  Name & {\it \rm lon [$^{\rm \circ}$]} & {\it \rm lat [$^{\rm \circ}$]} & {\rm d [${\rm kpc}$]} & {\it $ \rm  TS\footnote{TS = $ {\chi^2_{\rm w/oALP,min} - \chi^2_{\rm wALP,min}}$}_{Bfield1}$} \\[3pt]
\hline
IC443 &189.065 & 3.235 & 1.5  & 20.9 \\[3pt]
W51C & 49.131 & -0.467 & ${\rm 5.4 \pm 0.6}$  & 16.0 \\[3pt]
W49B & 43.275 & -0.190 & ${\rm 11.3 \pm 0.4}$ & 7.4 \\[3pt]
\end{tabular}
\end{center}
\end{ruledtabular}
\label{tb1}
\end{table}

\section{Photon-ALP Oscillation in The Milky Way Magnetic Field}\label{sec:2}

The $\gamma$-ray photons from SNRs may convert to ALPs (and vice versa) when passing through external magnetic fields. 
If the field strength is strong enough and the propagation distance is long enough, the cumulative effect of the photon-ALP oscillation can lead to a detectable modulation in the spectra of SNRs. 
The survival probability ${\rm P_{\rm ALP}}$ is intended to measure the probability of photons which have not been converted into ALPs at the end of the propagation along the lines of sight. 
Assuming a homogeneous magnetic field with the size $l$  and the transversal strength $B_{\rm T}$, the survival probability for an initially polarized photon with energy $E_\gamma$ can be approximately written as \cite{axionf,axionf1}
\begin{eqnarray}\label{eq:p}
P_{\rm ALP}&=&1-P_{\gamma \rightarrow a} \\ \nonumber
&=&1-\frac{1}{1+{E_{\rm c}^2}/{E_\gamma^2}} \sin^2 \left[\frac{g_{a\gamma} B_{\rm T} l}{2}  \sqrt{1+\frac{E_{\rm c}^2}{E_\gamma^2}} \right].
\end{eqnarray}
The mixing between photons and ALPs would be strong, when $E_\gamma$ is higher than the characteristic energy $E_{\rm c}$ defined as
\begin{eqnarray}
E_{\rm c}=\frac{\left| m_a^2-w_{\rm pl}^2\right|}{2g_{a\gamma} B_{\rm T}},
\end{eqnarray}
And $w_{\rm pl}^2=4\pi\alpha n_{\rm e}/m_{\rm e}$ represents the plasma frequency. 

In practice, we explicitly solve the evolution equation for photon-ALP beam as in Ref. \cite{axionf,axionf1} to calculate the survival probability, instead of using the simplified formulae Eq. (\ref{eq:p}).
For simplicity, we assume photons from the target sources are unpolarized in our analysis, which would yield conservative results.
Note that it has been shown that photon-photon dispersion can have important effects for extragalactic sources \cite{2017JCAP...01..024K} and it is especially relevant for the VHE range. In this work,  we also take into account this effect. However, it is found that for all the three Galactic sources considered in this work, the contribution from photon-photon dispersion $\Delta_{\gamma\gamma}$ is much less than the terms related to the ALP mass $\Delta_{a}$ and the photon-ALP coupling $\Delta_{a\gamma}$. Neglecting the photon-photon dispersion will in fact not affect the results.

The large-scale Galactic magnetic field model developed by Jansson \& Farrar \cite{Bfield1} (Bfield1) is adopted in our work, which has been wildly used in the relevant investigations \cite{Majumdar17psr1,ALP2018,Majumdar17psr2,ic443alp,liang18alp}.
This model composes of a disk field, an extended halo field and out of plane component and provides a best fit to WMAP7 Galactic synchrotron emission map and more than 40,000 extragalactic rotation measures (RM) data.
We update magnetic field parameters ($b_6^{\rm disc}=-3.5$, $B_X=1.8$) with the newly published data of the polarized synchrotron and dust emission from the Planck satellite \cite{2016A&A...596A.103P}. Unless specially mentioned, the results in our paper are based on Bfield1.
Apart from the large-scale regular component, the small-scale random field is also a reasonable part of the Galactic magnetic field.
Since its coherence length is much smaller than the photon-ALP oscillation length, we don't take it into account in this work.

The magnetic field strength and the thermal electron density of the Milky Way respectively are $B_{\rm T} \sim 1\,\mu{\rm G}$ and $n_{\rm e} \sim 0.1\,{\rm cm^{-3}}$. 
We found if $g_{a\gamma}=10 \times 10^{-11}\,{\rm GeV}^{-1}$ and $m_a=1\,{\rm neV}$, the characteristic energy is about 250 MeV, beyond which the HE and VHE observations have a high sensitivity. 
The parameter spaces we investigate cover the range of $g_{a\gamma}$ ($0.1-100 \times 10^{-11}\,{\rm GeV}^{-1}$) and $m_a$ ($0.1-1000\,{\rm neV}$).
Figure \ref{fig:PALP} illustrates the survival probabilities for photons from IC443, with five sets of parameters in the region we scan.

\section{ Fermi-LAT Spectrum}\label{sec:3}
The Large Area Telescope aboard the Fermi Gamma-Ray Space Telescope Mission (Fermi-LAT) is currently the most sensitive high-energy $\gamma$ -ray telescope in the energy range from below 100 MeV to over 300 GeV \cite{fermilat}.
Compared with previous versions, the {\tt Pass 8} Fermi-LAT data has an extension of the energy range, better energy measurements and a larger effective area \cite{pass8econf}.
In this work, we use nearly ten years of LAT {\tt Pass 8} (P8R3, the latest version) data from Oct. 27, 2008 to Aug. 8, 2018.
Before Oct. 27, 2008, the photon data have a significantly higher level of background contamination at energies above $\sim$ 30 GeV \footnote{\url {https://fermi.gsfc.nasa.gov/ssc/data/analysis/LAT_caveats.html}}. So we exclude the earlier data.

The photon events ranging from 300 MeV to 800 GeV in the {\tt SOURCE} event class with the {\tt FRONT+BACK} conversion type were selected in our analysis. 
We exclude those events coming from zenith angles larger than $90^{\rm \circ}$ to reduce the contribution from the Earth's limb and adopted the recommended quality-filter cuts {\tt(DATA\_QUAL==1 \&\& LAT\_CONFIG==1)} to extract the good time intervals. A $14^{\rm \circ} \times 14^{\rm \circ}$ Region Of Interest (ROI) binned in $0.1^{\rm \circ}$ spatial bins centered on each target SNR is defined for a standard binned likelihood analysis.
Our following analysis is performed using the updated Science Tools package {\tt Fermitools} \footnote{\url {https://github.com/fermi-lat/Fermitools-conda/}} and instrument response functions (IRFs) {\tt P8R3\_SOURCE\_V2} provided by the Fermi-LAT collaboration \footnote{\url {http://fermi.gsfc.nasa.gov/ssc/}} .

We use the user-contributed script {\tt make4FGLxml.py} \footnote{\url{http://fermi.gsfc.nasa.gov/ssc/data/analysis/user/}} to make an initial model for each SNR, which is a combination of the updated standard Galactic diffuse emission model (gll\_iem\_v07.fits), the updated standard isotropic diffuse model for the SOURCE data (iso\_P8R3\_SOURCE\_V2\_v1.txt) and all the fourth Fermi-LAT source catalog (4FGL) \cite{4fgl} sources located within $15^{\rm \circ}$ from the position of each SNR.
Based on eight years of Fermi-LAT data, the 4FGL has nearly twice of the source number of the third Fermi-LAT source catalog (3FGL) \cite{3fgl}.
In our analysis, the spectral shape of each source is set as the default setting in the 4FGL . 
Spectral indexes of sources within $5^{\rm \circ}$ of the target SNR and all the normalizations in ROI are considered to be free parameters. 
In the fitting procedure, we use the Fermi-LAT {\tt  pyLikelihood} code to get an optimized model.

To derive bin-by-bin flux points, we divide the data into 30 evenly logarithmically-spaced energy bins from 300 MeV to 800 GeV. 
In this step, we fix spectral indexes of all the sources to values optimized in the global fit. 
 And all the normalizations are set free to vary during the likelihood fit.
Then the standard binned likelihood analysis is performed in each energy bin. 
The Test Statistic (TS) value is defined as two times the logarithmic likelihood ratio between the signal hypothesis and the null hypothesis of the target source in each energy bin.
When the TS value of the target SNR is smaller than 25, the 95\% upper limit of the flux is computed in this energy bin by using the {\tt pyLikelihood UpperLimits} tool .

\section{Joint Analysis and Result}\label{sec:4}

To search for possible ALP-photon oscillation signals, we combine the spectra of Fermi-LAT and IACT observations into a joint broadband spectrum for each SNR.
Then we make the $\chi^2$ analysis by refitting obtained broadband spectra with the intrinsic spectrum model (the null model) and the ALP model separately, which has been widely applied in Ref. \cite{fermi13pi0, Jogler16w51c, Majumdar17psr1, Majumdar17psr2, ALP2018, ic443alp, liang18alp}.

The intrinsic spectra of all the targets are modeled by a {\tt LogParabola} function\footnote{\url {https://fermi.gsfc.nasa.gov/ssc/data/analysis/scitools/source\_models.html}}, the same as that adopted in the 4FGL \cite{4fgl}
\begin{equation}
{{\left(\frac{dN}{dE}\right)}_{\rm SNR}=N_0\left(\frac{E}{E_{\rm b}}\right)^{-[\alpha+\beta {\rm log}(E/E_{\rm b})]} },
\label{eq:LogParabola}
\end{equation}

The ALP model is the result of multiplying the intrinsic spectrum by the survival probability $P_{\rm ALP}$, which can be expressed as 
\begin{equation}
{{\left(\frac{dN}{dE}\right)}_{\rm wALP}=P_{\rm ALP}(g_{a\gamma}, m_{a}, E){\left(\frac{dN}{dE}\right)}_{\rm SNR}}.
\label{eq:ALP}
\end{equation}

To handle energy dispersion of instruments, we convolve both spectral models with the energy dispersion function $D_{\rm eff}(E',E)$
\begin{equation}
{\frac{dN}{dE'}}=D_{\rm eff}\otimes{\frac{dN}{dE}},
\label{eq:Deff}
\end{equation}
where $E'$ and $E$ represent the observed and the true energy, respectively.
The energy dispersion of Fermi-LAT and IACTs instruments are considered separately. 
For the Fermi-LAT,  we calculate the exposure weighted energy dispersion function $D_{\rm eff}$ by \cite{2016PhRvD..93j3525L},
\begin{equation}
D_{\rm eff}(E',E)=\frac{\sum_{\rm j}{\epsilon}(E',{\theta}_{\rm j}) {\cdot} D(E',E,{\theta}_{\rm j})}{\sum_{\rm j} {\epsilon}(E',{\theta}_{\rm j})},
\end{equation}
where $D$ is the energy dispersion function of Fermi-LAT from the FSSC website\footnote{\url{https://fermi.gsfc.nasa.gov/ssc/data/analysis/documentation/Cicerone/Cicerone_LAT_IRFs/IRF_E_dispersion.html}}, and ${\epsilon}$ is the exposure as a function of incline angle from the boresight $\theta$.
In the case of IACTs, the energy dispersion is approximated as a Gaussian function with the standard deviation $\sigma$ defined as the energy resolution of the instrument, similar to the Ref. \cite{liang18alp} . In the literature \cite{magicer16, veritas, veritaser17, hess_line1, hess_line2, hesser09}, the energy resolutions of IACTs are reported in the range of 10\% to 20\%.  However, the energy dispersion of the IACTs is related to the actual observation data and the detailed information is not fully available. But in the Ref. \cite{liang18alp}, it is found that the inaccurate use of the energy resolution (within 10\% to 20\% range) has only minor effects on the final results. The exact values of the IACT energy resolutions we use in the analysis will be stated in the below specific subsection for each SNR.

To investigate the signature of photon-ALP oscillations, we then use the above null (ALP) model to refit the broadband spectra to obtain the value of $\chi^2_{\rm w/oALP}$ ($\chi^2_{\rm wALP}$). 
We derive a TS (defined as TS = ${\chi^2_{\rm w/oALP} - \chi^2_{\rm wALP}}$) map by repeated this procedure for a grid of ALP masses $m_{a}$ and photon-ALP couplings $g_{a\gamma}$ covering the range of $(0.1-100) \times 10^{-11}\,{\rm GeV}^{-1}$ and $0.1-1000\,{\rm neV}$, respectively.
Besides, in the case of either model, $N_0$, $\alpha$, $\beta$ and $E_{\rm b}$ are left free during fitting procedures.
After scanning a grid of ($m_{a}$, $g_{a\gamma}$), we obtain optimal ALP parameters or exclude inappropriate parameter spaces.

\subsection{IC443}\label{sec:4a}

\begin{figure*}
\centering
\includegraphics[width=0.49\textwidth]{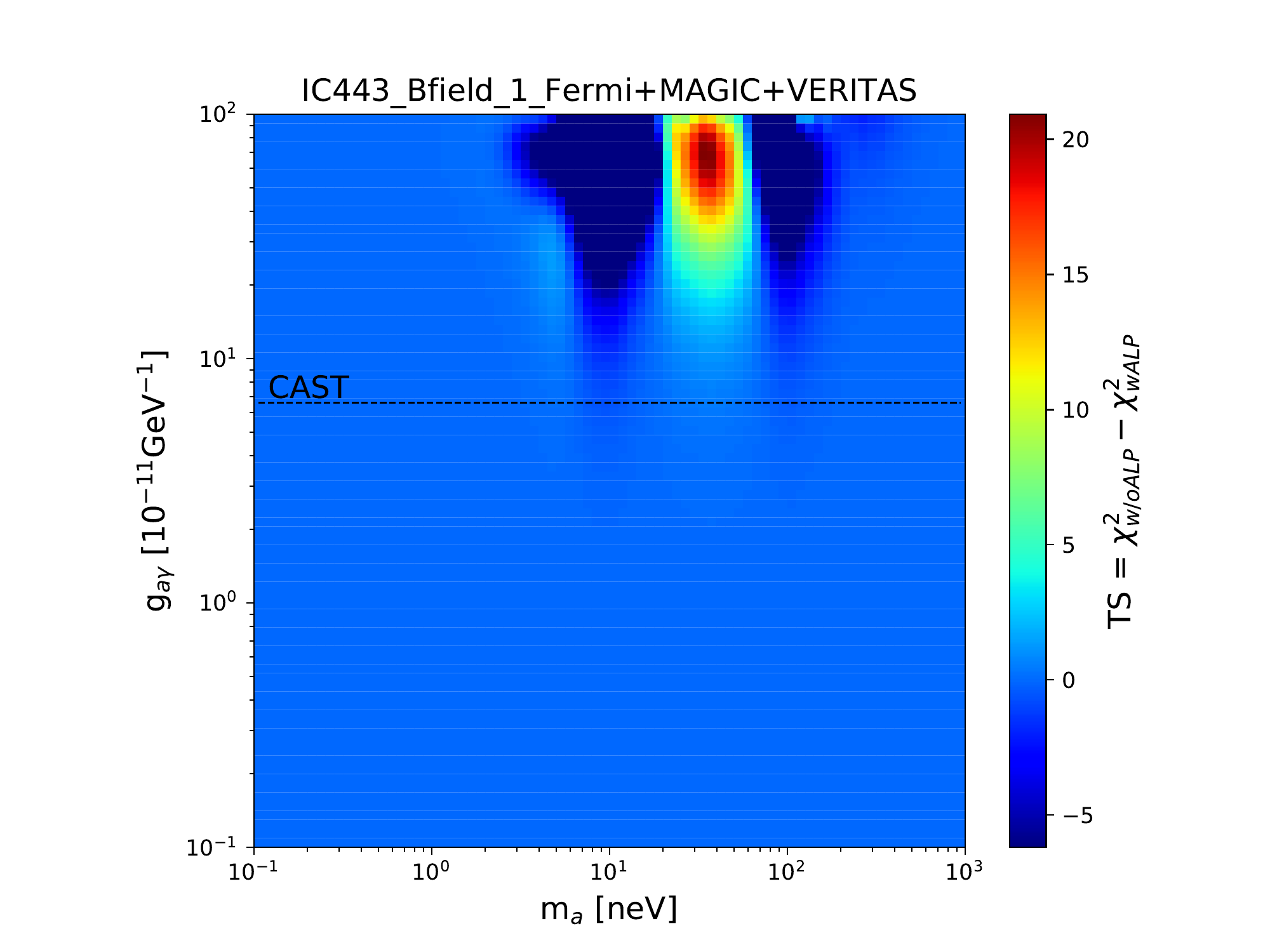}
\includegraphics[width=0.49\textwidth]{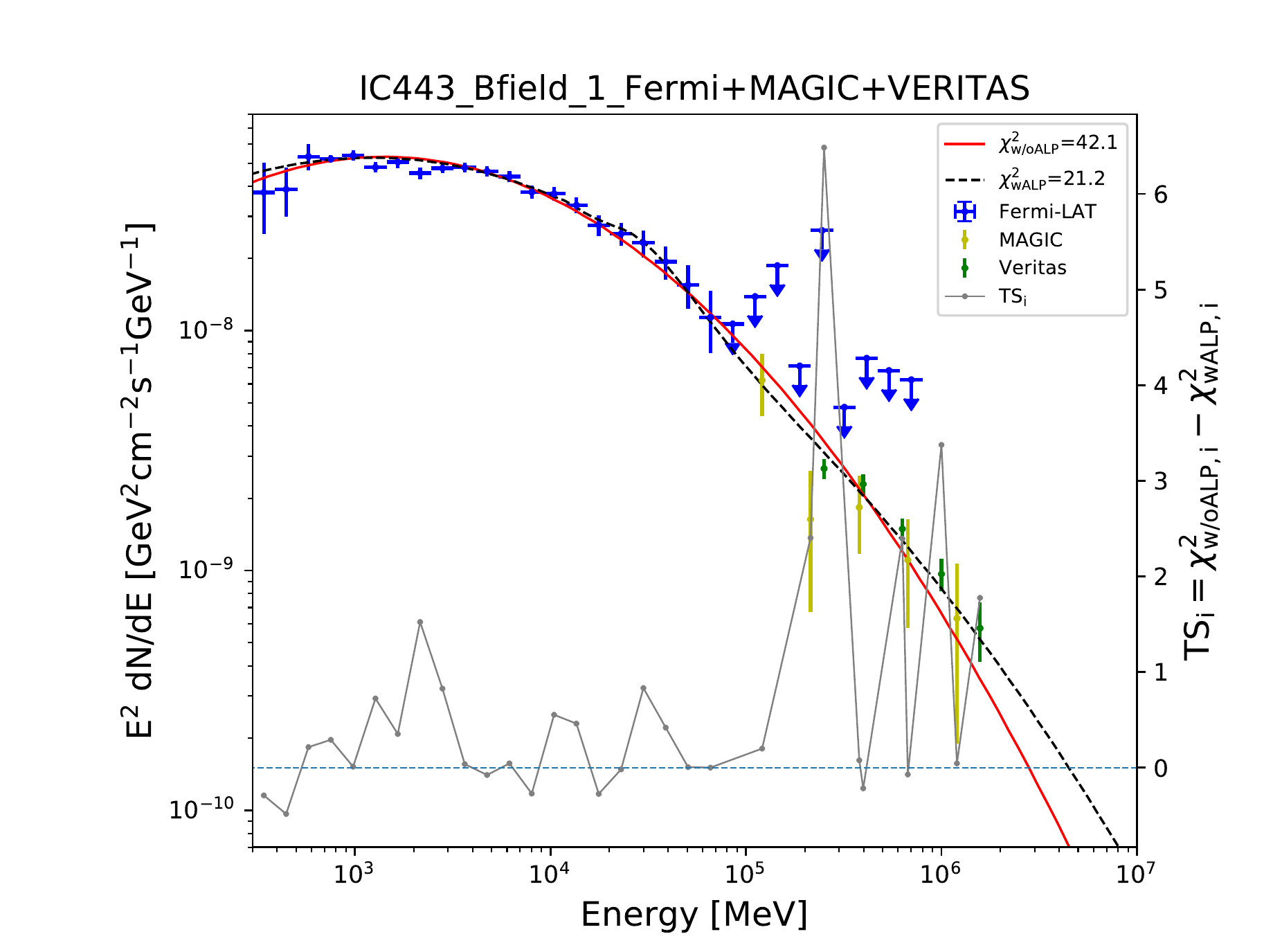}
\caption{Left panel: The TS value as a function of ALP mass $m_{a}$ and photon-ALP coupling constant $g_{a\gamma}$ for IC443 (Fermi+MAGIC+VERITAS, Bfield1).
The upper boundary of the color bar scale is set as the TS$_{\rm max}$ in the scanning, while the lower one is $-{\rm 6.2}$. 
The black dashed lines show the upper limit of the photon-ALP coupling set by CAST \cite{CAST2017}. 
Right panel: The joint broadband spectrum and the best-fit models without and with photon-ALP oscillations.
The ${\rm TS_i} = \chi^2_{\rm w/oALP, i} - \chi^2_{\rm wALP, i}$ (gray curve) demonstrates the contribution of each data point to the total TS value for the best-fit ALP model.}
\label{fig:IC443}
\end{figure*}

The TeV emission from IC443 has been detected by MAGIC \cite{ic443magic07} and VERITAS \cite{ic443veritas09, ic443veritas15}.
The observations with MAGIC were carried out in the periods December 2005 - January 2006 (10 hours) and December 2006 - January 2007 (37 hours) \cite{ic443magic07}. 
The energy-dependent energy resolution of the MAGIC telescope given by Ref. \cite {magicer16} was adopted in our analysis, which is approximately 15\%-20\% from 100 GeV to 15 TeV. 
VERITAS observations of IC443 began in February, 2007, and concluded in January, 2015, with an exposure of approximately 155 hours live time \cite{ic443veritas15}. 
The energy resolution of VERITAS reported in the literature ranges from 10\% to 20\% \cite{veritas,veritaser17} (17\% at 1 TeV\footnote{\url{https://veritas.sao.arizona.edu/about-veritas-mainmenu-81/veritas-specifications-mainmenu-111}}).
For conservativeness, we adopt a value of 20\% in our work. 

We combine the Fermi-LAT spectrum of IC443 with the spectra observed by IACTs (MAGIC \cite{ic443magic07} and VERITAS \cite{ic443veritas15}).
 The comparisons between the models and the observations are given in the right panel of Figure \ref{fig:IC443}. The TS values of the two-dimensional parameter space ($m_{a}$, $g_{a\gamma}$) are shown in the left panel of Figure \ref{fig:IC443}, where the red (blue) regions are favored (disfavored) by the joint Fermi+MAGIC+VERITAS spectra.

As we can see in Figure \ref{fig:IC443}, in the case of Bfield1, IC443 gives a relatively high TS value $\sim20.9$ of the ALP model with the best-fit ALP mass and coupling constant ($m_{a}$, $g_{a\gamma}$) = ($33.5~{\rm neV}$, $67.8 \times 10^{-11}~{\rm GeV}^{-1}$). The corresponding confidence level is 3.24 $\sigma$, following the $\chi^2$ distribution with 4.6 degrees of freedom (see details in Section \ref{sec: 4}).

However these best-fit parameters are in tension with the current upper limits set by the globular clusters and the CAST experiment (the black dashed line in the left panel of Figure \ref{fig:IC443}). In addition, the high TS value is mainly contributed by the 
VERITAS data (see the grey line in the right panel of Figure \ref{fig:IC443}).
 It may be due to the different absolute energy calibration between VERITAS and the Fermi-LAT, which will be discussed in detail in Section \ref{sec: ec}. It is noted that we only consider statistical uncertainties of the VHE data in our analysis. As shown in Ref. \cite{2009ApJ...698L.133A}, systematic uncertainties on the integral flux and the photon index of IC443 (VERITAS) are similar with the corresponding statistical uncertainties. This may be a possible reason for the mismatch between the Fermi-LAT extrapolation and the VHE spectra.


The current result is inconsistent with our pervious work \cite {ic443alp} which solely analyzed the Fermi-LAT spectrum of IC443 with the P8R2 data and the 3FGL catalog \cite{3fgl}. Our pervious work \cite {ic443alp} gave the TS value $\sim20.8$ ($4.2\sigma$) at the favored parameters ($m_{a}$, $g_{a\gamma}$) = ($6.7~{\rm neV}$, $20.2 \times 10^{-11}~{\rm GeV}^{-1}$) for Bfield1. Such parameters however are strongly excluded by the current data. This may be due to the changes in the Fermi-LAT spectrum of IC443 derived using the P8R3 data and 4FGL (instead of the P8R2 data and 3FGL), which we will further examine in Section \ref{sec: fl}.

\begin{figure*}
\centering
\includegraphics[width=0.49\textwidth]{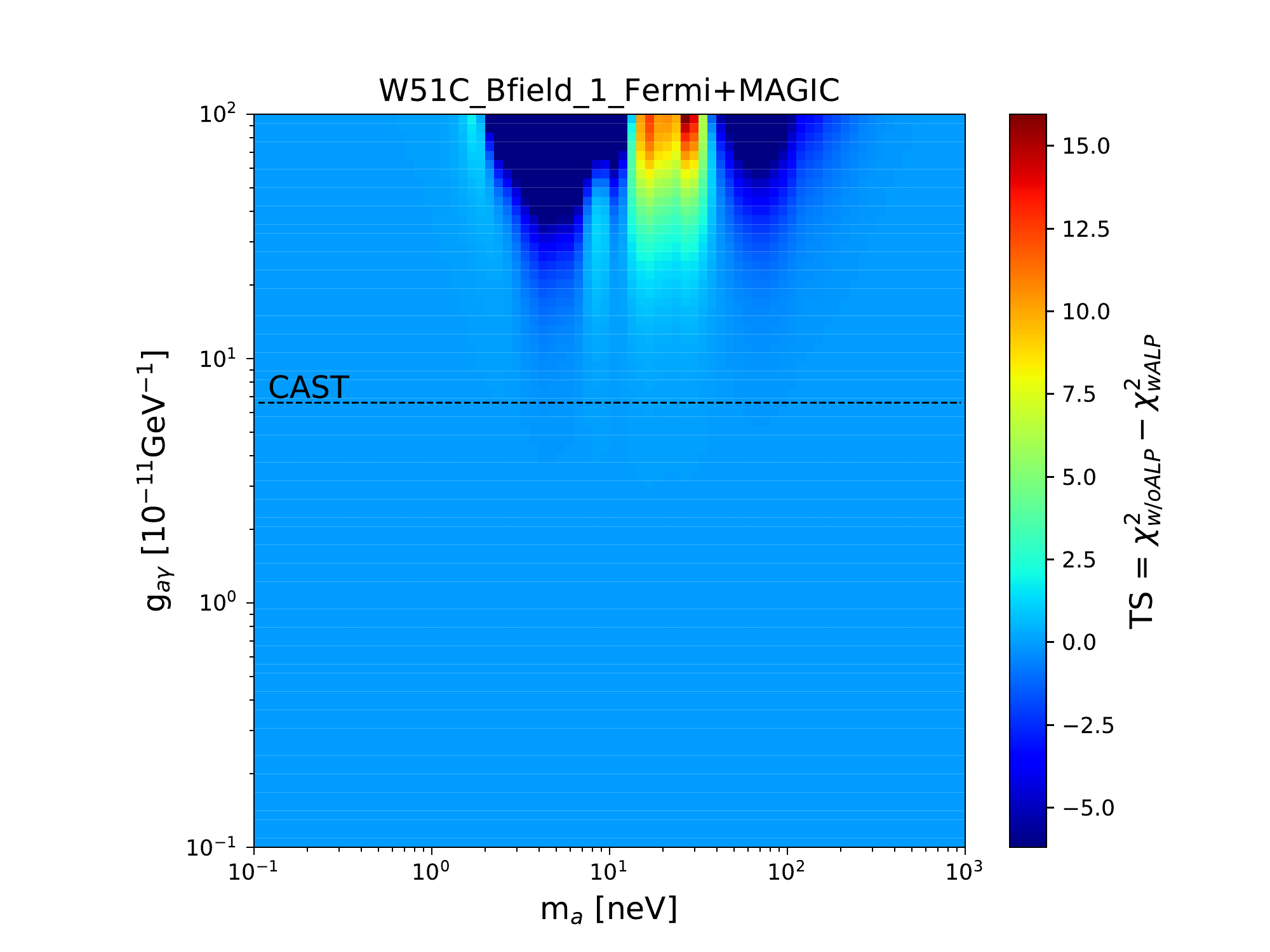}
\includegraphics[width=0.49\textwidth]{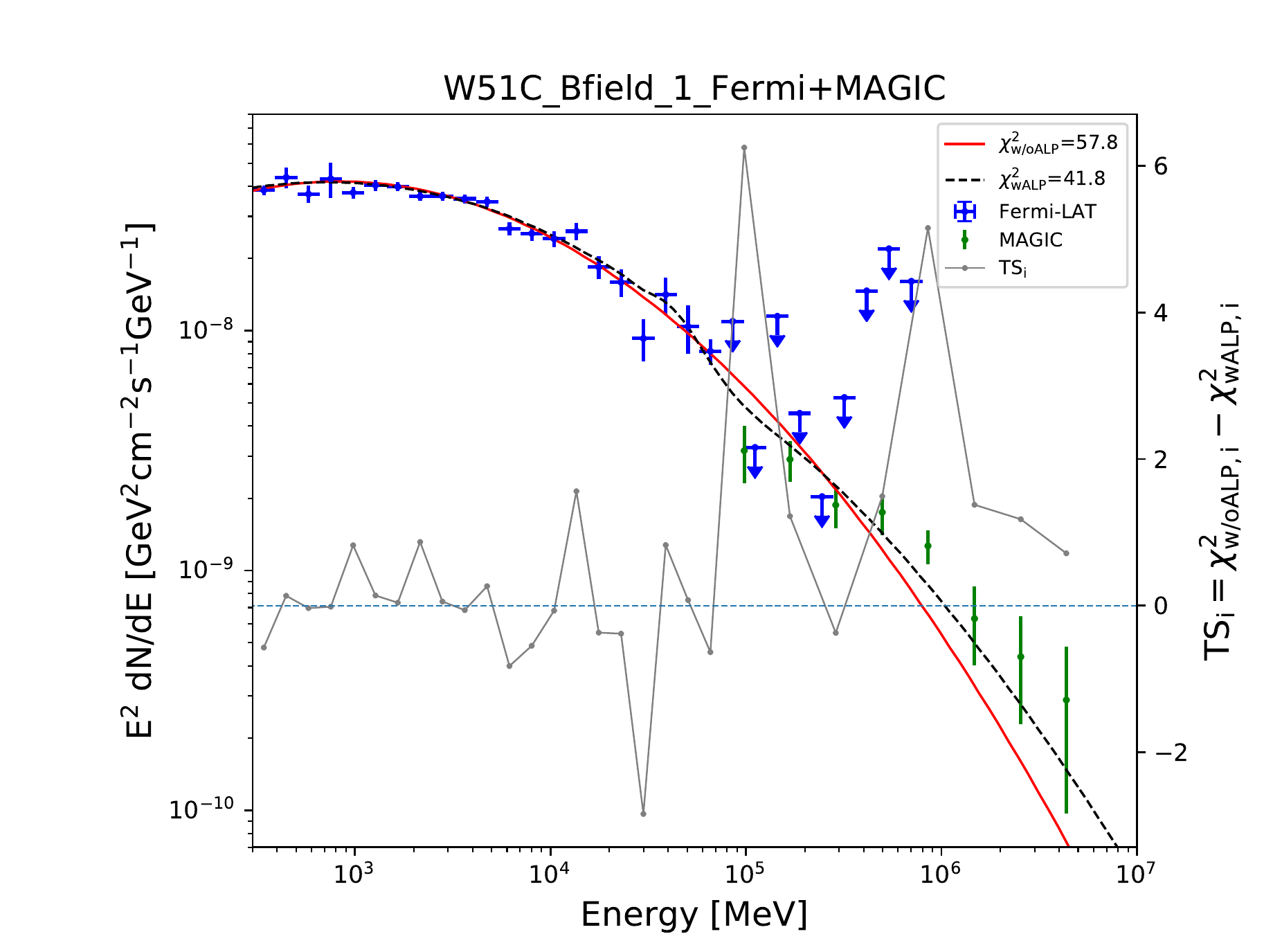}
\caption{Left panel: The TS value as a function of ALP mass $m_{a}$ and photon-ALP coupling constant $g_{a\gamma}$ for W51C (Fermi+MAGIC, Bfield1).
Right panel: The joint broadband spectrum of W51C and the best-fit models without and with photon-ALP oscillations.}
\label{fig:W51C}
\end{figure*}


\subsection{W51C}
We use the MAGIC spectrum of W51C derived by a data set comprising 53 hours effective time between 2010 and 2011 \cite{w51cmagic12}. 
The energy resolution of MAGIC is taken as the same as IC443.
The results of the joint Fermi+MAGIC analysis for W51C, given in Figure \ref{fig:W51C}, imply that W51C prefers the ALP model of the best-fit parameters ($m_{a}$, $g_{a\gamma}$) = ($26.6~{\rm neV}$, $95.8 \times 10^{-11}~{\rm GeV}^{-1}$) over the null model with the TS value $\sim 15.9$ for Bfield1.

However, as shown in the left panel of Figure \ref{fig:W51C}, the best-fit parameters have been also excluded by the CAST limit, like IC443. As we can see in the right panel of Figure \ref{fig:W51C}, the TS value $\sim 15.9$ is predominantly contributed by the MAGIC spectrum, at a confidence level of 2.56 $\sigma$. The same as IC443, the credibility of the high TS value ($\sim$15.9) of the ``signal" in W51C is also uncertain, since we could not exclude that it is from different absolute energy calibration between MAGIC and the Fermi-LAT, which we will discuss in Section \ref{sec: ec}.

\begin{figure*}
\centering
\includegraphics[width=0.49\textwidth]{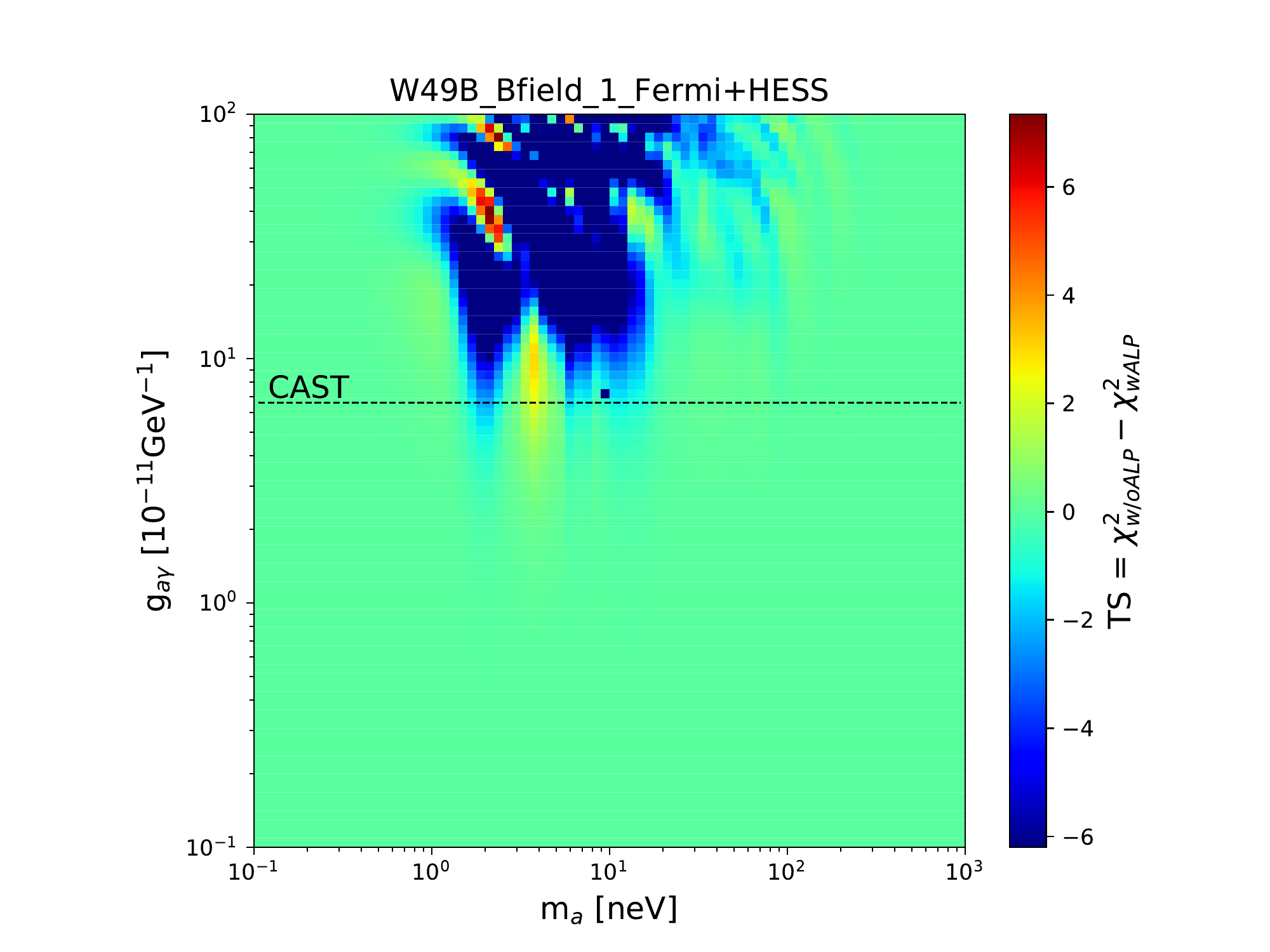}
\includegraphics[width=0.49\textwidth]{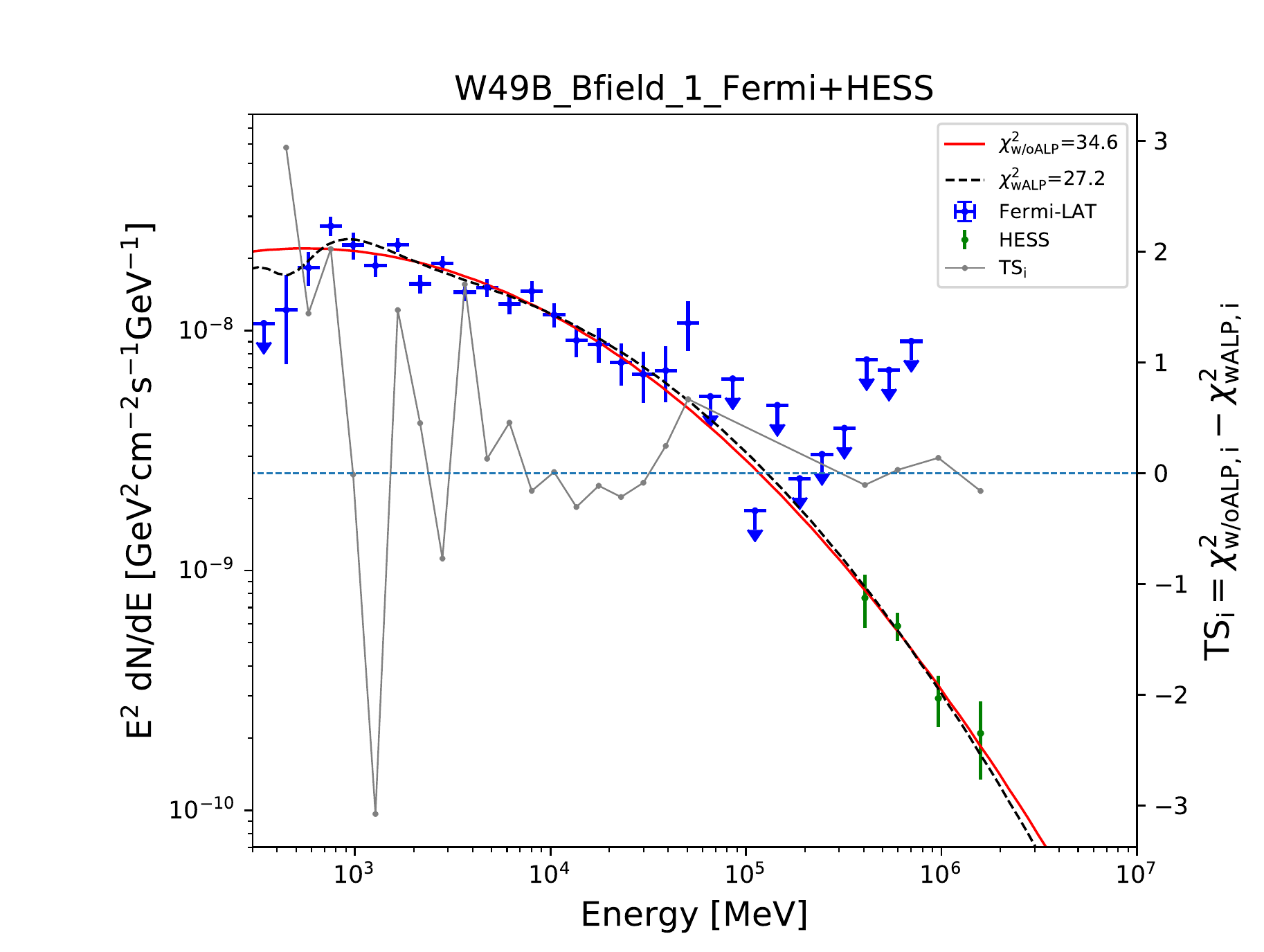}
\caption{Left panel: The TS value as a function of ALP mass $m_{a}$ and photon-ALP coupling constant $g_{a\gamma}$ for W49B (Fermi+\hess, Bfield1). 
Right panel: The joint broadband spectrum of W49B and the best-fit models without and with photon-ALP oscillations.}
\label{fig:W49B}
\end{figure*}

\subsection{W49B}
\hess observed W49B between 2004 and 2013 in a total amount of 75 hours live time \cite{w49bhess18}.
The energy resolution of \hess ranges from 10\% to 20\% \cite{hess_line1, hess_line2, hesser09}.
We use the energy-dependent energy resolution presented in Ref. \cite{hesser09} to do the joint Fermi+\hess analysis.
The result is plotted in the Figure \ref{fig:W49B}. No significant preference with TS $>$ 10.5 (corresponding to a confidence level of 2 $\sigma$, see details in Section \ref{sec: 4} )  of the ALP model is found.

\section{Combined limits of all the three SNRs}
\label{sec: 4}

Since all the tentative signals are still doubtful, in the following, we set constraints on ALP parameters with the combined analysis of all the three SNRs.
We define $\Delta\chi^2=\chi_{\rm wALP}^2-\chi_{\rm w/oALP}^2$ to quantify how the ALP model is disfavored by the observation data.

By summing the $\Delta\chi^2$ value of all the three SNRs, we derive combined constraints on ALP parameters. 
Some studies \cite{fermi16alp, liang18alp} have shown that, due to the complexity between ALP parameters and spectral irregularities, the $\Delta\chi^2$ distribution in the null hypothesis is not as expected by Wilks’ theorem \cite{wilks1938} to be a $\chi^2$ distribution with 2 degree of freedom. 
We thus perform Monte Carlo (MC) simulation to derive the null distribution and the threshold value of $\Delta\chi^2$ .
For this purpose, we first generate the pseudo spectra of Fermi-LAT and VHE.

In the case of Fermi-LAT, we use the optimized model obtained in the global fit of Section \ref{sec:3} to make a model map
with the binning and size the same as the observed counts cube (CCube). Based on the obtained model map, MC simulated CCubes are generated following a Poisson statistic.  To be more specified, the number of events in each pixel is determined assuming a Poisson distribution with its mean $\lambda$ the model-expected counts in the corresponding pixel.
We perform the same likelihood analysis as in Section \ref{sec:3} on these simulated CCubes to derive the flux points of the MC spectra.
In order to speed up the calculation, we fix all background parameters to the best-fit values in the fit of simulation data.

For the IACTs data, since the IRFs and exact exposure information are not fully available, we use the same strategy as in the Ref. \cite{liang18alp} to create the pseudo spectra.
For each flux point in the observed spectrum, the nominal flux value of the simulated spectrum is randomly assigned based on a Gaussian distribution.
The mean of the Gaussian is from the best-fit null model and the sigma is the uncertainty of the observation. The error bars in the simulated spectrum are set to the same values as the observed one. 

\begin{figure}
\centering
\includegraphics[width=0.5\textwidth]{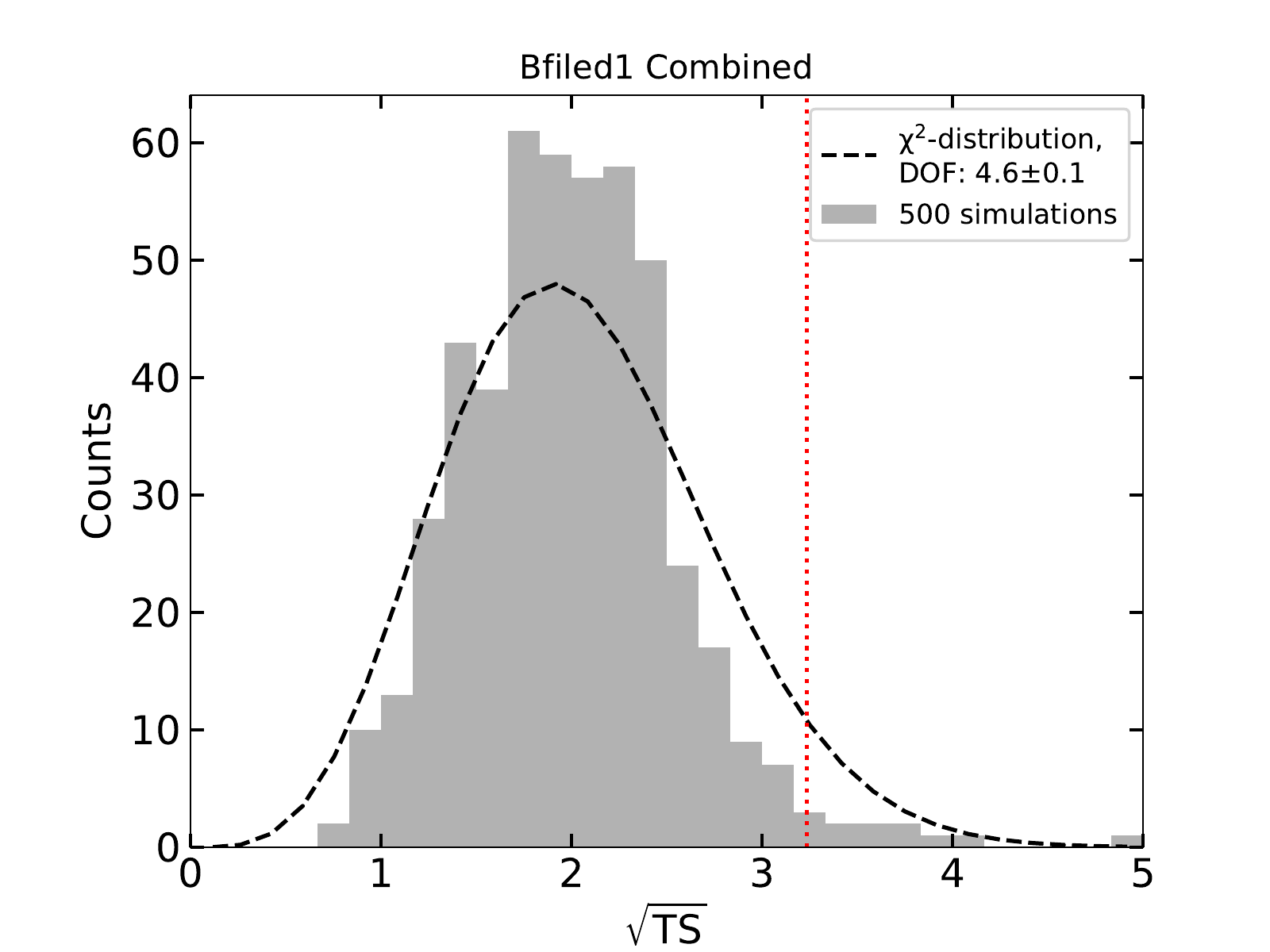}
\caption{The TS distribution of the null hypothesis for the ALP analysis. The histogram is from 500 Monte-Carlo simulations of spectra with the null hypothesis. The dashed curve is the best-fit $\chi^2$ distribution with degrees of freedom $d = 4.6\pm 0.1$. The red vertical line shows the threshold value corresponding to 95\% confidence level based on the best-fit distribution.}
\label{fig:dis}
\end{figure}

After preparing the pseudo spectra, we apply the same ALP analysis as used in Section \ref{sec:4} to them to scan two-dimensional parameter space ($m_a$, $g_{a\gamma}$). For each source, we carry out 500 simulations. The TS distribution for 500 Monte-Carlo simulations is shown in Figure \ref{fig:dis}. Using a non-central $\chi^2$ function to fit the histogram yields degrees of freedom $d=4.6\pm0.1$ and non-centrality parameter $s = 0$. Though the $\chi^2$ distribution in fact can not fit the real distribution very well, the latter is narrower than the former, thus using the best-fit $\chi^2$ to determine the threshold value would lead to conservative results. Based on this distribution, we can derive the threshold value $\Delta\chi^2_{thr}$=10.5 above which ALP parameters will be excluded at 95\% confidence level. 

\begin{figure}
\centering
\includegraphics[width=0.5\textwidth]{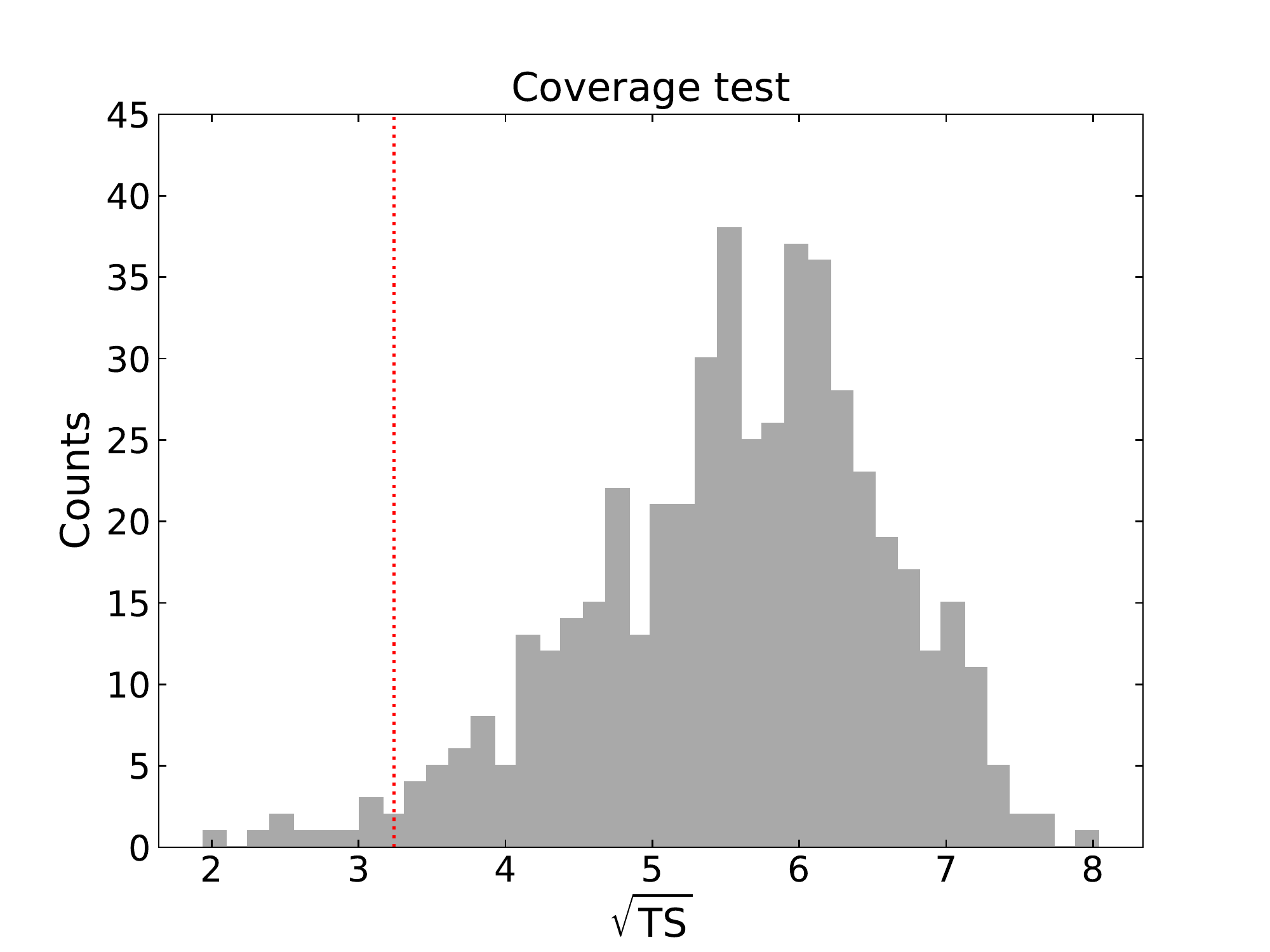}
\caption{The TS distribution for the coverage test with 500 simulations of spectra with the injected ALP signal ($m_{a}$, $g_{a\gamma}$) = ($10~{\rm neV}$, $50 \times 10^{-11}~{\rm GeV}^{-1}$).  }
\label{fig:ct}
\end{figure}

To make a coverage test, we also simulate spectra of all the three SNRs with an ALP signal for 500 times and see how often the injected ALP signal can be recovered. Considering the computational cost, we choose a subset of ALP parameters ($m_{a}$, $g_{a\gamma}$) = ($10~{\rm neV}$, $50 \times 10^{-11}~{\rm GeV}^{-1}$) which is excluded by our combined analysis. Then we find the injected ALP signal can be recovered by 484 times with TS$>$10.5 in 500 simulations. Figure \ref{fig:ct} shows the TS distribution for this coverage test.

\begin{figure}
\centering
\includegraphics[width=0.52\textwidth]{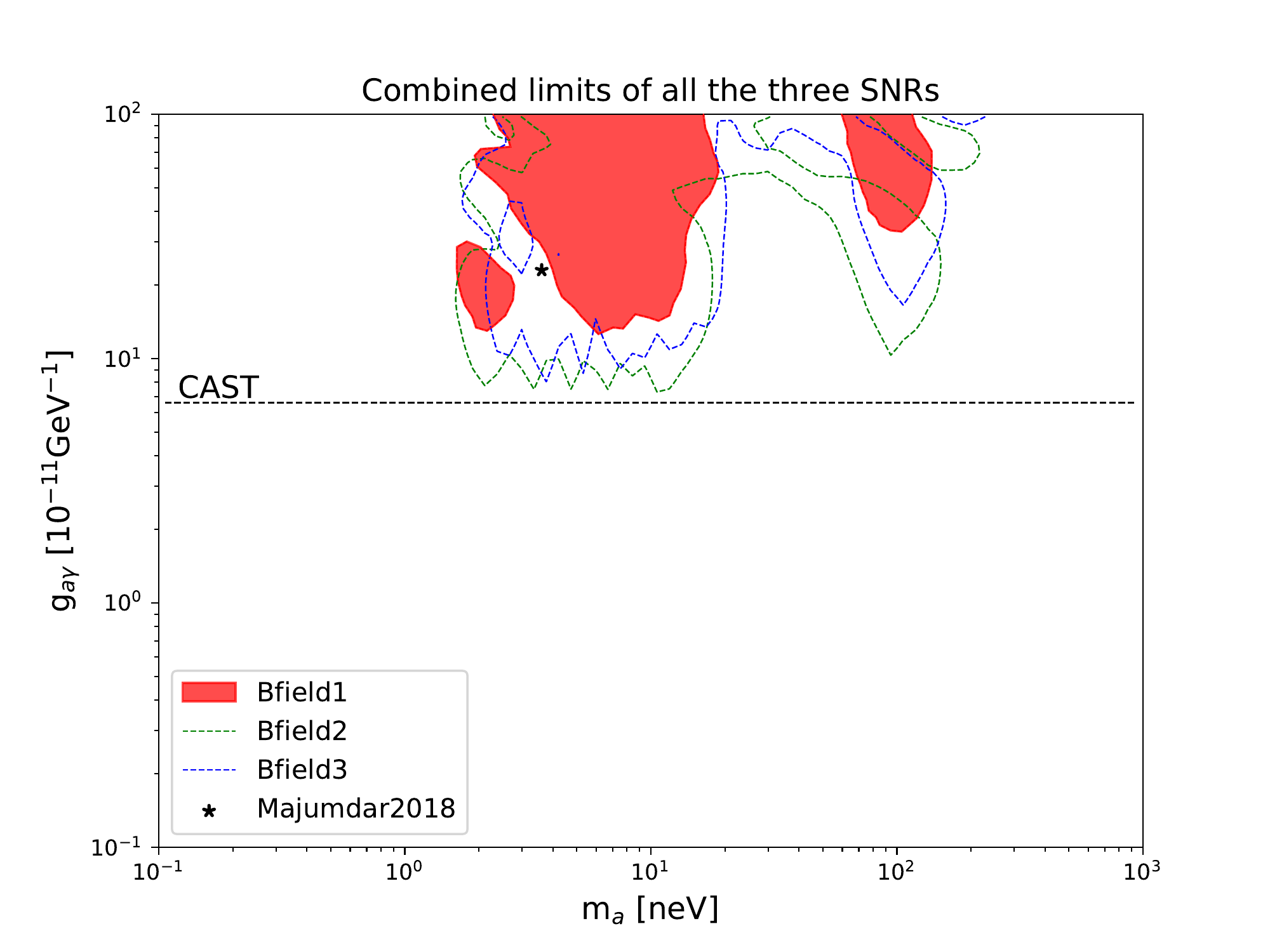}
\caption{The 95\% confidence level limits on ALP parameters by combining all the three SNRs. The red region is obtained from Bfield1. The dashed lines are for two additional magnetic field models Bfield2 (green) and Bfield3 (blue), respectively. The black star ($m_{a}$, $g_{a\gamma}$) = ($3.6~{\rm neV}$, $23\times 10^{-11}~{\rm GeV}^{-1}$) is the best-fit ALP parameters of the possible photon-ALPs mixing signal found by Majumdar et al. with the Jansson and Farrar (Bfield1) GMF model in Ref. \cite{ALP2018}.}
\label{fig:cl}
\end{figure}

The results of the combined limits are showed in Figure \ref{fig:cl}. The red region is constrained at 95\% confidence level for Bfield1. Compared with the CAST limits (the black dashed line in Figure \ref{fig:cl}), the combined analysis of all the three SNRs don't give tighter constraints, which however could be regarded as a supplement and complement to the pervious limits.

\section{Discussion}\label{sec:5}

\subsection{The Galactic magnetic field model}
The interactions between ALPs and photons need the participation of the Milky Way magnetic field.
Therefore, the choice of the Galactic magnetic field model could induce the uncertainty in our analysis.

In order to test this uncertainty, we repeat the analysis with two additional Galactic magnetic field models which are developed by Sun et al. \cite{Bfield2} (Bfield2) and Pshirkov et al. \cite{Bfield3} (Bfield3), respectively.
Compared with the Bfield1, the Bfield2 models the magnetic field in the disk and the halo of the Galaxy by the use of multi-wavelength synchrotron and RM data \cite{Bfield2}. For the Bfield3, Pshirkov et al. constrain the model consisting of two components (disk and halo) with two sets of RM data of extragalactic radio source \cite{Bfield3}.

Though stronger than those for Bfield1, the combined limits with Bfield2 (green, dashed) and Bfield3 (blue, dashed) shown in Figure \ref{fig:cl} are also weaker than the limits set by the CAST experiment. However, so far we haven't known enough about the Milky Way magnetic field and the GMF models used in our work are still not perfect. Further improvements, such as better knowledge of the local environment \cite{Bfield1} and more extragalactic RM data \cite{Bfield3}, are expected in the future.

\subsection{The Fermi-LAT spectrum of IC443}
\label{sec: fl}

In our pervious work \cite {ic443alp}, the Fermi-LAT spectrum of IC443 obtained with the P8R2 data and 3FGL showed a preference for the ALP model at the best-fit parameters ($m_{a}$, $g_{a\gamma}$) = ($6.7~{\rm neV}$, $20.2 \times 10^{-11}~{\rm GeV}^{-1}$) corresponding to a TS value $\sim20.8$ (Bfield1).
However, with these best-fit parameters, the ALP model don't give any improvement to the fit of the joint Fermi+MAGIC+VERITAS spectrum of IC443 in this work.

\begin{figure*}
\centering
\includegraphics[width=0.49\textwidth]{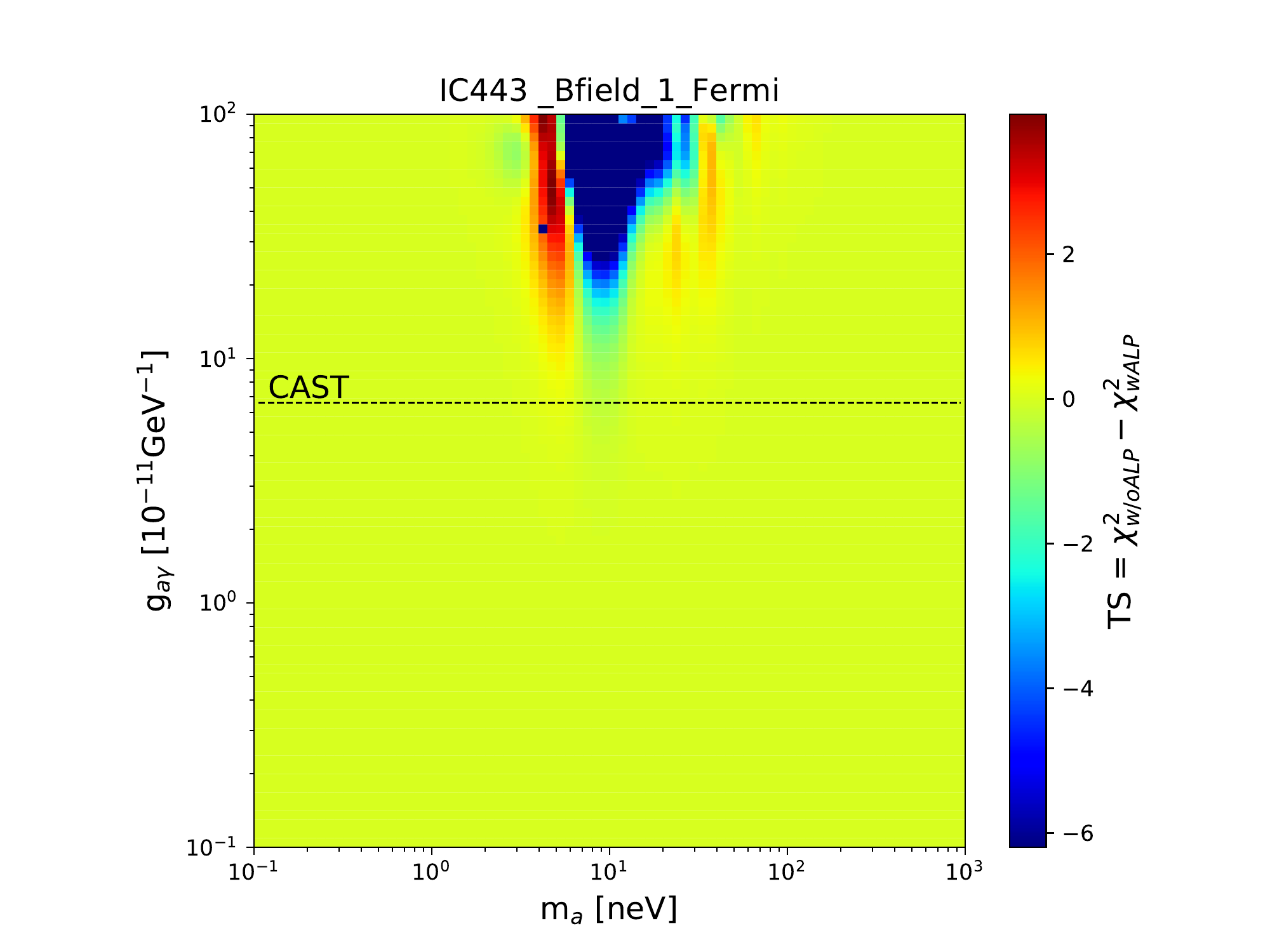}
\includegraphics[width=0.49\textwidth]{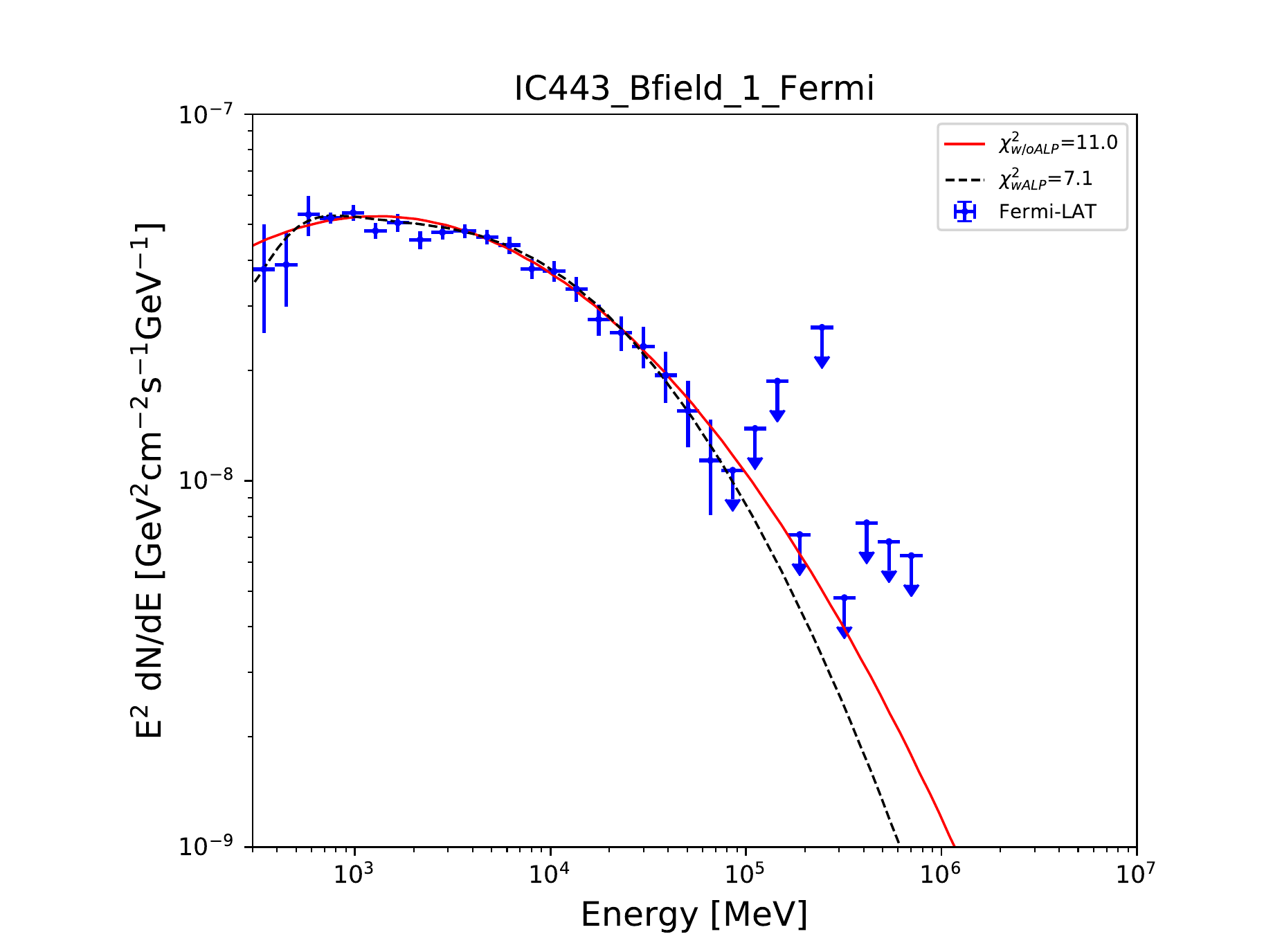}
\caption{Left panel: The TS value as a function of ALP mass $m_{a}$ and photon-ALP coupling constant $g_{a\gamma}$ for IC443 (Fermi, Bfield1).
Right panel: The Fermi-LAT spectrum of IC443 and the best-fit models without and with photon-ALP oscillations.}
\label{fig:ICF}
\end{figure*}

We wonder if such distinct results are induced by the addition of IACTs data in the analysis or the different versions of Fermi-LAT data (P8R2/P8R3) and model files (3FGL/4FGL). So in the following, we use the new Fermi-LAT spectrum of IC443 solely to repeat the analysis in the pervious work \cite {ic443alp}.
The results, shown in Figure \ref{fig:ICF}, indicate that the ALP model isn't obvious better than the null model in the parameter spaces we investigate for the new Fermi-LAT spectrum and the best-fit TS value is 3.9, which are inconsistent with our previous work\cite {ic443alp}. These results indicate that the different ALP results in the two works are mainly due to the different Fermi-LAT spectra.

Compared with our pervious work \cite {ic443alp}, this work use the latest version of the Fermi-LAT data (P8R3) and the model file with the updated diffuse emission templates (gll\_iem\_v07.fits, iso\_P8R3\_SOURCE\_V2\_v1.txt) and the new Fermi-LAT source catalog (4FGL) to get the Fermi-LAT measured spectrum. It is noteworthy that there are two new point sources within $0.5^{\rm \circ}$ from the center of IC443 in 4FGL. They are 4FGL J0616.5+2235 ($0.16^{\rm \circ}$ from the center of IC443 ) and 4FGL J0618.9+2240 ($0.43^{\rm \circ}$, which is associated with 3FGL J0619.4+2242 but the location has changed). In the 4FGL, the newly added point source 4FGL J0616.5+2235 is classified as ``spp" which represents sources of unknown nature but overlapping with the location of a known SNR or PWN. In our analysis, we don't take 4FGL J0616.5+2235 as a part of IC443, but in the 3FGL the flux/spectrum of IC443 contains its contribution.

Besides, in the Fermi-LAT only analysis, we also find that TS value decrease to 0.27 at the joint-analysis favored parameters ($m_{a}$, $g_{a\gamma}$) = ($33.5~{\rm neV}$, $67.8 \times 10^{-11}~{\rm GeV}^{-1}$) of IC443. It seems that the high TS value we get in the joint analysis of IC443 is mainly induced by the unsmooth connection between the Fermi-LAT spectrum and the IACT observations, which could be as a result of the different absolute energy calibrations of the instruments (see in detail in Section \ref{sec: ec}).

\subsection{Absolute energy calibration}
\label{sec: ec}

\begin{table}[t]
\caption{Energy scaling factors for IACTs}
\begin{ruledtabular}
\begin{center}
\begin{tabular}{ccccc}
Source & IACT & Scaling Factor ($S_{\rm factor}$) & $ {\chi^2_{\rm w/oSfactor}}$/d.o.f.& $ {\chi^2_{\rm wSfactor}}$/d.o.f. \\[3pt]
\hline
IC443 & MAGIC & ${\rm 1.44 \pm 0.001}$  & 1.10 & 0.69\\[3pt]
IC443 & VERITAS & ${\rm 1.47 \pm 0.0006}$ &  1.77 & 0.92\\[3pt]
\hline 
W51C & MAGIC &  ${\rm 1.24 \pm 0.0007}$  & 2.51 & 2.15 \\[3pt]
\hline 
W49B & \hess & ${\rm 1.16 \pm 0.002}$ & 2.05 & 2.01 \\[3pt]
\end{tabular}
\end{center}
\end{ruledtabular}
\label{tb2}
\end{table}

One of the main systematic uncertainties in our analysis is the different absolute energy calibrations between the Fermi-LAT and IACTs.
In order to estimate this effect, we assume the energy scaling factor of Fermi-LAT to be 1 and use the joint Fermi-LAT and IACT spectra to derive scaling factors ($S_{\rm factor}$) for each IACT.
To obtain the common energy scale $E $ of IACT, we use the measured energy $E_{\rm meas}$ to multiply the energy scaling factor $S_{\rm factor}$ of each IACT \cite{2010A&A...523A...2M}:
\begin{equation}
E=E_{\rm meas}\cdot S_{\rm factor}
\label{eq:Deff}
\end{equation}
Then we use the null (intrinsic) model to fit the joint Fermi-LAT and modified IACT data and the energy scaling factors ($S_{\rm factor}$) of IACT is left as a free parameter in the fitting processes.

The resulting scaling factors for IACTs are listed in Table \ref{tb2} together with the reduced minimum ${\chi^2}$ values without and with scaling factors in the fits.
However, the IACT energy scaling factors we obtain are larger than those reported in Ref. \cite{2010A&A...523A...2M} by the analysis of Crab Nebula (5-10\%). One possible reason for the difference is that the intrinsic spectra are simply modeled by the {\tt LogParabola} function in our analysis. While in Ref.\cite{2010A&A...523A...2M}, the predicted inverse compton component is used as the intrinsic model in the HE and VHE band to derive energy scaling factors for IACTs. 
A better understanding of the astrophysical processes with the multi-wavelengths observations is to be expected in the further analysis. As mentioned in Section \ref{sec:4a}, we have not taken systematic uncertainties of the VHE data into account, which could also explain large IACTs energy scaling factors we get.

Then we introduced these energy scale factors into the joint analysis of IC443. Corresponding to ALP parameters ($m_{a}$, $g_{a\gamma}$) = ($23.7~{\rm neV}$, $37.1 \times 10^{-11}~{\rm GeV}^{-1}$), the best-fit TS value of the ALP signal reduces to 4.8, which is too weak to be a reliable signal.
Future telescopes with significantly improved performance may reduce this uncertainty related to the absolute energy calibration, which would make the results of the joint HE+VHE analysis more instructive.

\section{Sumary}\label{sec:6}
In this work, we search for the possible signal of the photon-ALP oscillation using the joint Fermi-LAT and IACTs spectra of three SNRs. 
In the case of IC443 and W51C, the joint spectra give relatively high TS values of 20.9 and 15.9, respectively, for the model including the photon-ALP oscillation. However the corresponding best-fit ALP parameters are in tension with the CAST limits. Moreover, we find these high TS values may be due to unsmooth transitions between Fermi-LAT and IACTs. We only consider the statistical uncertainties of IACTs data in our analysis. And we speculate that these unsmooth transitions might be introduced by systematic uncertainties of the absolute energy calibrations of instruments. In Section \ref{sec: ec} , we derive energy scaling factors of instruments using the joint Fermi-LAT and IACTs spectra.

Since the tentative signals found in our analysis are still in doubt, we constrain ALP parameters using the combined analysis of all the three SNRs. Except Bfield1, we take two more additional GMF models (Bfield2 and Bfield3) into account to set the corresponding limits. 
For all three magnetic field models we choose, the combined limits are weaker than the limits set by the CAST experiment and observations of globular clusters.
Regardless, these combined limits we get provide a complementary support to the existing constraints. 
Based on the Bfield1 GMF model, Ref. \cite{ALP2018} found the possible indication of the photon-ALP oscillation in the $\gamma$-ray spectra of bright pulsars. The corresponding best-fit ALP parameters are ($m_{a}$, $g_{a\gamma}$) = ($3.6~{\rm neV}$, $23\times 10^{-11}~{\rm GeV}^{-1}$). It is interesting to note that our combined limits couldn't exclude this hint found by Ref. \cite{ALP2018} with the same GMF model (Bfield1) (see Figure \ref{fig:cl}).

For the joint analysis of IC443, we obtain a different result compared to our pervious work only using the Fermi-LAT spectrum of IC443 \cite{ic443alp}. 
The difference likely arises from the using of the latest version of the Fermi-LAT data (P8R3), the new model file with the updated diffuse emission model template (gll\_iem\_v07.fits, iso\_P8R3\_SOURCE\_V2\_v1.txt) and the new Fermi-LAT source catalog (4FGL) .

Note that there are some other telescopes sensitive to the GeV or TeV energy range at present or in the future, which may contribute significantly to the search of photon-ALP oscillations. 
In particular, the Cherenkov Telescope Array (CTA), the next generation IACT, will cover the energy range from a few tens of GeV to above 100 TeV with the improved sensitivity, enhanced angular and energy resolutions over existing VHE $\gamma$-ray observatories \cite{cta}.
Besides, the Dark Matter Particle Explorer (DAMPE) also displays outstanding energy resolution in a wide energy range of 10 GeV to TeV and above \cite{DAMPE1,DAMPE2}.
With more and more $\gamma$-ray telescopes joining, the existence of ALP will be investigated more deeply in the near future.

\begin{acknowledgments}
This work is supported by the National Key Research and Development Program of China (Grant No. 2016YFA0400200), the National Natural Science Foundation of China (Grants No. 11525313, No. 11722328, No. 11773075, No. U1738210, No. U1738136, No. 11773075 and No. U1738206), the 100 Talents Program of Chinese Academy of Sciences, and the Youth Innovation Promotion Association of Chinese Academy of Sciences (Grant No. 2016288).
\end{acknowledgments}

\bibliographystyle{apsrev4-1-lyf}
\bibliography{alp_GTSNR}

\end{document}